# Heating or Cooling: Study of Advective Heat Transport in the Inflow and the Outflow of Optically Thin Advection-dominated Accretion Flows

Cheng-Liang Jiao (焦承亮)[1,2,3]
[1] Yunnan Observatories, Chinese Academy of Sciences, 396 Yangfangwang, Guandu District, Kunming, 650216, People's Republic of China; jiaocl@ynao.ac.cn
[2] Center for Astronomical Mega-Science, Chinese Academy of Sciences, 20A Datun Road, Chaoyang District, Beijing, 100012, People's Republic of China
[3] Key Laboratory for the Structure and Evolution of Celestial Objects, Chinese Academy of Sciences, 396 Yangfangwang, Guandu District, Kunming, 650216,
People's Republic of China

## Abstract

Advection is believed to be the dominant cooling mechanism in optically thin advection-dominated accretion flows (ADAFs). When outflow is considered, however, the first impression is that advection should be of opposite sign in the inflow and the outflow, due to the opposite direction of radial motion. Then how is the energy balance achieved simultaneously? We investigate the problem in this paper, analyzing the profiles of different components of advection with self-similar solutions of ADAFs in spherical coordinates ($r\theta\phi$). We find that for $n < 3\gamma/2 - 1$, where $n$ is the density index in $\rho \propto r^{-n}$ and $\gamma$ is the heat capacity ratio, the radial advection is a heating mechanism in the inflow and a cooling mechanism in the outflow. It becomes 0 for $n = 3\gamma/2 - 1$, and turns to a cooling mechanism in the inflow and a heating mechanism in the outflow for $n > 3\gamma/2 - 1$. The energy conservation is only achieved when the latitudinal ($\theta$ direction) advection is considered, which takes an appropriate value to maintain energy balance, so that the overall effect of advection, no matter the parameter choices, is always a cooling mechanism that cancels out the viscous heating everywhere. For the extreme case of $n = 3/2$, latitudinal motion stops, viscous heating is balanced solely by radial advection, and no outflow is developed.

*Unified Astronomy Thesaurus concepts:* Accretion (14); Hydrodynamics (1963); High energy astrophysics (739)

## 1. Introduction

Heat transport by advective motion is a very important mechanism in accretion disk theory, which has given rise to models distinct from the standard thin disk (Shakura & Sunyaev 1973), such as the optically thin advection-dominated accretion flow (ADAF; e.g., Narayan & Yi 1994, 1995a, 1995b; Abramowicz et al. 1995; see Yuan & Narayan 2014, for a review) and the slim disk (e.g., Abramowicz et al. 1988; Chen & Taam 1993; Jiao et al. 2009). In these models, advection usually acts as a cooling mechanism (hence the terminology "advective cooling"), carrying away the leftover energy of the local viscous heating that is not radiated away due to inefficiency of radiation (for ADAFs) or a large optical depth (for slim disks). Theoretically speaking, advection is an effective heating/cooling mechanism when we focus on a fixed region in the rest frame, which is essentially the net difference in entropy between the incoming and the outgoing material for this specific region (Kato et al. 2008). Apparently it could be a heating mechanism equally, depending on properties of the flow motion, and a natural prediction is that it should be of opposite sign in the inflow and the outflow, due to the opposite direction of radial motion.

Recent observations have provided some evidence of strong outflow from ADAFs. Wang et al. (2013) proposed the existence of strong outflow from the ADAF onto the supermassive black hole in the Galactic center, based on Chandra observation of the quiescent X-ray emission of Sgr A*. Park et al. (2019) studied Faraday rotation in the jet of M87 obtained from Very Long Baseline Array data sets, which are well described by a gas density profile $\rho \propto r^{-1}$, consistent with an ADAF with substantial winds. Shi et al. (2021, 2022) reported the detection of winds from ADAFs in the low-luminosity active galactic nuclei in M81 and NGC 7213, based on high-resolution Chandra observations.

There have also been many theoretical and numerical simulation works concerning outflow in ADAFs. Narayan & Yi (1994, 1995a) mentioned the possibility of outflow due to positive Bernoulli parameters in the original ADAF model. Stone et al. (1999) performed two-dimensional hydrodynamic (HD) numerical simulation of ADAFs and found that strong outflow is generated while only a small fraction of the accreted material reaches the central black hole, which is later confirmed by both HD and magnetohydrodynamic (MHD) simulations (e.g., Yuan et al. 2012 and references therein; Li et al. 2013; Yuan et al. 2015; Yang et al. 2021). Two competing models were proposed to explain this phenomenon. The adiabatic inflow–outflow solution (ADIOS; Blandford & Begelman 1999, 2004; Begelman 2012) features strong outflow that carries away mass, angular momentum, and energy from the inflow. On the other hand, in the convection-dominated accretion flow (CDAF; Narayan et al. 2000; Quataert & Gruzinov 2000), the majority of accreted material is locked in turbulent convective eddies arising from convective instability, so that the net accretion rate remains small while none of the accreted material really escapes (but could be inaccurately counted as outflow in numerical simulations depending on the integration method; see Yuan & Narayan 2014 for a detailed description). Recent numerical simulations found that ADAFs are actually convectively stable (e.g., Yuan et al. 2012; Narayan et al. 2012). Yuan et al. (2015) (and later Yang et al. 2021 for Kerr black holes) further investigated the trajectories of virtual Lagrangian particles from simulation data, and found that there exists strong "real" outflow in the form of disk wind and jet, while the wind form dominates the mass outflow rate. Under convective stability, local viscous heating in an ADAF can no longer be cooled by convection, nor by radiation due to its inefficiency, leaving advection as the only possible dominant cooling mechanism. Then a natural question is how advection acts as a cooling mechanism simultaneously in both the inflow and the outflow, which have opposite radial motion. We investigate the question in this paper. Decomposing advective heat transport based on direction (radial and latitudinal) and mechanism (advected internal energy and pressure work), we study how each component works in the inflow and the outflow. The basic equations and assumptions are presented in Section 2. The formulae of advection components are analyzed in Section 3, and their profiles are presented and discussed in Section 4. We summarize and conclude in Section 5.

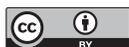

## 2. Basic Equations and Assumptions

The model that we use to analyze advective heat transport is basically the same as presented in Jiao & Wu (2011), and we focus on ADAFs here. We consider a steady ($\partial/\partial t = 0$) and axisymmetric ($\partial/\partial \phi = 0$) ADAF onto a compact object at the origin in spherical coordinates ($r\theta\phi$), neglecting the self-gravity of accreted material. The equations of continuity and motion can be respectively written as (Kato et al. 2008)

$$\nabla \cdot (\rho \boldsymbol{v}) = 0, \tag{1}$$

$$\rho \boldsymbol{v} \cdot \nabla \boldsymbol{v} = -\rho \nabla \Psi - \nabla p + \nabla \cdot \boldsymbol{T}, \tag{2}$$

where $\Psi$ is the gravitational potential and $\boldsymbol{T}$ is the tensor of viscous stress. Here we adopt the Newtonian gravitational potential, $\Psi = -GM/r$. We also assume that the $r\phi$-component of the viscous stress tensor, $t_{r\phi}$, is dominant in the accretion flow, and adopt the $\alpha$ prescription of viscosity (Shakura & Sunyaev 1973), $t_{r\phi} = -\alpha p$.

The energy equation is written as

$$\nabla \cdot (E \boldsymbol{v}) + p \nabla \cdot \boldsymbol{v} = f q_{\text{vis}}, \tag{3}$$

where the left-hand side (LHS) represents the entropy change per unit volume, i.e., the advective heat transport $q_{\text{adv}}$, $E$ is the internal energy density per unit volume, $q_{\text{vis}}$ is the heating rate of viscous dissipation, which is

$$q_{\text{vis}} = t_{r\phi} r \frac{\partial}{\partial r}\left(\frac{v_\phi}{r}\right), \tag{4}$$

when only the $r\phi$-component of $\boldsymbol{T}$ is considered, and $f$ is the advective factor. For ADAFs, we expect $f \to 1$, i.e., the local viscous heating needs to be balanced by the advective heat transport completely for a steady solution. Due to radiative inefficiency, here we treat the accretion flow as gas pressure dominated, so that $p = p_{\text{gas}}$, and $E = p/(\gamma - 1)$, where $\gamma = c_p/c_v$ is the heat capacity ratio. In ADAFs, temperature is high, usually close to virial temperature, so we would expect $\gamma \to 5/3$, the value for monoatomic ideal gas. Note that if radiation and/or magnetic energy are considered, $\gamma$ should be replaced with an equivalent value $\gamma_e = p/E + 1$, which is lower than 5/3. The value of $\gamma$ will also be lower when relativistic effects are considered, becoming 4/3 for an extreme relativistic ideal gas, though generally the nonrelativistic value should be a good enough approximation (see, e.g., Esin 1997; Quataert & Narayan 1999; Yuan & Narayan 2014, for more discussions on $\gamma$). The formula of $q_{\text{adv}}$ used here is equivalent to the form used in other related papers (e.g., Narayan & Yi 1995a; Xue & Wang 2005; Jiao & Wu 2011) when the continuity equation is taken into account, and we use this form because it clearly shows that $q_{\text{adv}}$ can be divided into two parts: the change in internal energy due to advective motion (the first term) and the pressure work (the second term).

We seek self-similar solutions of the above equations in the form

$$\rho = \rho(\theta) r^{-n}, \tag{5}$$

$$v_r = v_r(\theta) \sqrt{\frac{GM}{r}}, \tag{6}$$

$$v_\theta = v_\theta(\theta) \sqrt{\frac{GM}{r}}, \tag{7}$$

$$v_\phi = v_\phi(\theta) \sqrt{\frac{GM}{r}}, \tag{8}$$

$$p = p(\theta) GM r^{-n-1}, \tag{9}$$

so that Equations (1)–(3) can be reduced to a set of ordinary differential equations (ODEs), which can then be numerically solved with proper boundary conditions. Numerical simulations (Yuan et al. 2012, 2012) have shown that within $10 r_s$ ($r_s = 2GM/c^2$ is the Schwarzschild radius) of the central compact object, the relativistic effects become strong and self-similarity no longer holds. Some studies (Narayan & Yi 1994; Narayan et al. 1997; Jiao et al. 2015) proposed that self-similarity is a good approximation in the range of intermediate radii away from the inner and the outer boundaries. So here we keep the analyses beyond $10 r_g$ and focus on the range where self-similarity behaves well. Newtonian potential is also a good approximation of the relativistic one in this range.

For boundary conditions, we assume that the accretion flow has reflection symmetry about the equatorial plane, so that

$$\theta = \frac{\pi}{2}: v_\theta = \frac{\partial \rho}{\partial \theta} = \frac{\partial p}{\partial \theta} = \frac{\partial v_r}{\partial \theta} = \frac{\partial v_\phi}{\partial \theta} = 0, \tag{10}$$

where only four conditions are independent. We set $\rho(\pi/2) = 1$ as the last boundary condition.

## 3. Analyses of Advection Formulae

Substituting Equations (5)–(9) into LHS of Equation (3), we get the formula of advective heat transport

$$q_{\text{adv}} = \frac{c_1}{\gamma - 1}\left\{\left(\frac{3}{2}\gamma - n - 1\right)v_r(\theta) + \gamma v_\theta'(\theta) + v_\theta(\theta)\right. \\ \left. \times \left[\gamma \cot(\theta) + \frac{\partial \ln p(\theta)}{\partial \theta}\right]\right\}, \tag{11}$$

where

$$c_1 = p\Omega_K = p(\theta) GM r^{-n-1} \sqrt{\frac{GM}{r^3}}. \tag{12}$$

To investigate the difference of $q_{\text{adv}}$ in the inflow and the outflow, we divide it into two parts, the radial advection $q_r$, and the latitudinal advection $q_\theta$ (the azimuthal advection is 0 because of axisymmetry), each of which is then further divided into an advected internal energy term (with a subscript 1) and a pressure-work term (with a subscript 2):

$$q_r = q_{r1} + q_{r2} = c_1 \frac{(3\gamma - 2n - 2)v_r(\theta)}{2(\gamma - 1)}, \quad (13)$$

$$q_{r1} = c_1 \frac{(1 - 2n)v_r(\theta)}{2(\gamma - 1)}, \quad (14)$$

$$q_{r2} = c_1 \frac{3v_r(\theta)}{2}, \quad (15)$$

$$q_\theta = q_{\theta 1} + q_{\theta 2} = \frac{c_1}{\gamma - 1} \times \left\{ \gamma v_\theta'(\theta) + v_\theta(\theta) \left[ \gamma \cot(\theta) + \frac{\partial \ln p(\theta)}{\partial \theta} \right] \right\}, \quad (16)$$

$$q_{\theta 1} = \frac{c_1}{\gamma - 1} \{ v_\theta'(\theta) + v_\theta(\theta) [\cot \theta + \frac{\partial \ln p(\theta)}{\partial \theta} ] \}, \quad (17)$$

$$q_{\theta 2} = c_1 [v_\theta'(\theta) + \cot \theta v_\theta(\theta)]. \quad (18)$$

To avoid possible confusion, here we clarify the physical meanings of the signs of important quantities in this paper. First, according to Equation (3), $q_{\text{adv}} > 0$ represents a cooling mechanism, while $q_{\text{adv}} < 0$ a heating mechanism. This also applies to components of $q_{\text{adv}}$, such as $q_r$ and $q_\theta$. Second, we consider material going toward the compact object at origin as inflow, with $v_r(\theta) < 0$, and material going away from the compact object as outflow, with $v_r(\theta) > 0$. We always have $c_1 > 0$, which does not influence the signs of the components of advective heat transport.

We can see that all the components are proportional to $(GM)^{3/2}r^{-n-5/2}$ and their values can be calculated once the latitudinal profiles are determined. As mentioned before, we have $\gamma \to 5/3$ for highly ionized gas, so we need to know the range of $n$ additionally to analyze the above formulae. Under self-similarity, the inflow mass accretion rate is $\dot{M}_{\text{in}} \propto r^s$, with $s = 3/2 - n$ according to Equations (5) and (6). So generally we expect $n \leqslant 3/2$. The accretion flow for $n = 3/2$ consists of pure inflow, which will be discussed in detail in Section 4.2. Here we focus on the $n < 3/2$ case, for which mass inflow rate decreases as radius decreases and outflow is developed. Numerical solutions (see Jiao & Wu 2011 for details) show that for $n < 3/2$, $v_\theta(\theta)$ is always negative due to latitudinal motion except for the equatorial value $v_\theta(\pi/2) = 0$.

For radial advection, first we notice that the sign of $q_{r2}$ is the same as $v_r(\theta)$, so it is always a heating mechanism in the inflow and a cooling mechanism in the outflow. The reason is that geometrical properties of the spherical coordinates result in the intersections of radial motion scaling with $r^2$, so with $v_r \propto r^{-1/2}$, more material comes in than that goes out for a fixed region in the inflow, i.e., material undergoes compression and is thus heated. Similarly, material undergoes expansion in the outflow and is thus cooled. For the advected internal energy $q_{r1}$, under self-similarity we have $E \propto r^{-n-1}$, so that for $n < 1/2$, there is more internal energy incoming than outgoing in the inflow, heating the fixed region. Similarly, $q_{r1}$ acts as a cooling mechanism in the outflow. The situation reverses for $n > 1/2$. For $n = 1/2$, we have $q_{r1} = 0$, which means that the amount of incoming and outgoing internal energy in the radial direction just cancels out each other. The total radial advection $q_r$ is a heating mechanism for $n < 3\gamma/2 - 1$ and a cooling mechanism for $n > 3\gamma/2 - 1$ in the inflow, and the situation reverses in the outflow.

The latitudinal advection is more complicated, as the formulae contain $v_\theta(\theta)$, $v_\theta'(\theta)$, and $\partial \ln p(\theta)/\partial \theta$. On the equator where $v_\theta(\theta) = 0$ and $v_\theta'(\theta) > 0$, we have both $q_{\theta 1} > 0$ and $q_{\theta 2} > 0$, so it is always a cooling mechanism. This cooling can offset the heating caused by viscous dissipation and sometimes radial advection (for $n < 3\gamma/2 - 1$), so that a steady solution can be achieved. Away from the equator, $q_\theta$ continues to take the form that balances the energy equation.

Ideally, the value of $n$ for a typical ADAF ($f = 1$) should be determinable from other input parameters, namely $\alpha$ and $\gamma$. This is beyond the scope of the paper, and we resort to other works in literature to limit the range of $n$. Theoretical studies of Begelman (2012) proposed that $s = 1$ for radiatively inefficient accretion flows and $s \lesssim 1$ with small radiative losses (note that their $n$ is our $s$), corresponding to $n \gtrsim 0.5$. However, we think that their result of $n = 0.5$ for ADAFs is debatable. As will be shown in Section 4.1, for $n \neq 3/2$, the solution requires either non-self-similar structure around the polar axis, or the net accretion rate (the sum of mass inflow and outflow rates) being 0. In the former case, entropy is transported from the outflow to the non-self-similar region, which is not considered in Begelman (2012). In the latter case, Mosallanezhad et al. (2021) showed that an additional cooling mechanism (such as thermal conduction) is required to obtain a solution. Note that the outgoing entropy or the additional cooling mechanism is also an energy loss, which effectively plays the same role as radiative loss in the analysis of Begelman (2012), so even for ADAFs with negligible radiation, we can still only get $n \gtrsim 0.5$. And while we expect that a single numerical simulation should give a definite value of $n$, numerical simulations of ADAFs performed by different groups, with similar or even the same input parameters, tend to have slightly different values of $n$. Yuan et al. (2012) reviewed earlier simulations of ADAFs and proposed $n \in [0.5, 1]$. Later simulations basically agree with this range, e.g., $n = 0.6$ away from the outer boundary in Li et al. (2013) and $s = 1$ in Yuan et al. (2015). So we expect $n \in [0.5, 1]$ for real accretion flows. In this range, contrary to the stereotype of "advective cooling," radial advection is actually a heating mechanism in the inflow because pressure work dominates over advection of internal energy, and a steady state in the inflow can only be achieved when latitudinal advection is considered.

## 4. Profiles of Advection Components

### 4.1. The Inflow–Outflow Case for n < 3/2

Numerical solutions of Equations (1)–(3) for $n < 3/2$ are presented and investigated in this subsection. We first discuss the likely situation of ADAFs, and set $\alpha = 0.1$, $\gamma = 5/3$, and $f = 1$. We choose $n = 0.5$ and 0.8 as two examples, which fall in the range of [0.5, 1], as proposed by theoretical and numerical simulation works (see Section 3).

The solutions are displayed in Figure 1, with blue solid lines and black dashed lines corresponding to the solutions for $n = 0.5$ and 0.8, respectively. Except for the latitudinal profiles of three velocity components, pressure, and density, we also show the latitudinal profile of isothermal sound speed $c_s = (p/\rho)^{1/2}$ in the bottom right panel. All the velocity components and sound speed are normalized by the Keplerian velocity $v_K = (GM/r)^{1/2}$. The density profile is effectively normalized by the density on the equatorial plane $\rho_{eq}$, as we set $\rho(\pi/2) = 1$, and consequently the pressure profile is normalized by $\rho_{eq} v_K^2$. We can see that the accretion flow consists of an inflow region around the equatorial plane and an outflow region beyond the inflow, with $v_r = 0$ on the boundary. Each calculation stops at some value of $\theta$ as it approaches the polar axis, when the density and the pressure become very small. The interpretation is that self-similarity is not applicable to the whole space and there exists a non-self-similar structure around the polar axis. Here we provide a brief explanation (see Jiao & Wu 2011 for details). The continuity equation can be expanded to

$$\frac{1}{r^2}\frac{\partial}{\partial r}(r^2 \rho v_r) + \frac{1}{r \sin\theta}\frac{\partial}{\partial \theta}(\sin\theta \rho v_\theta) = 0. \quad (19)$$

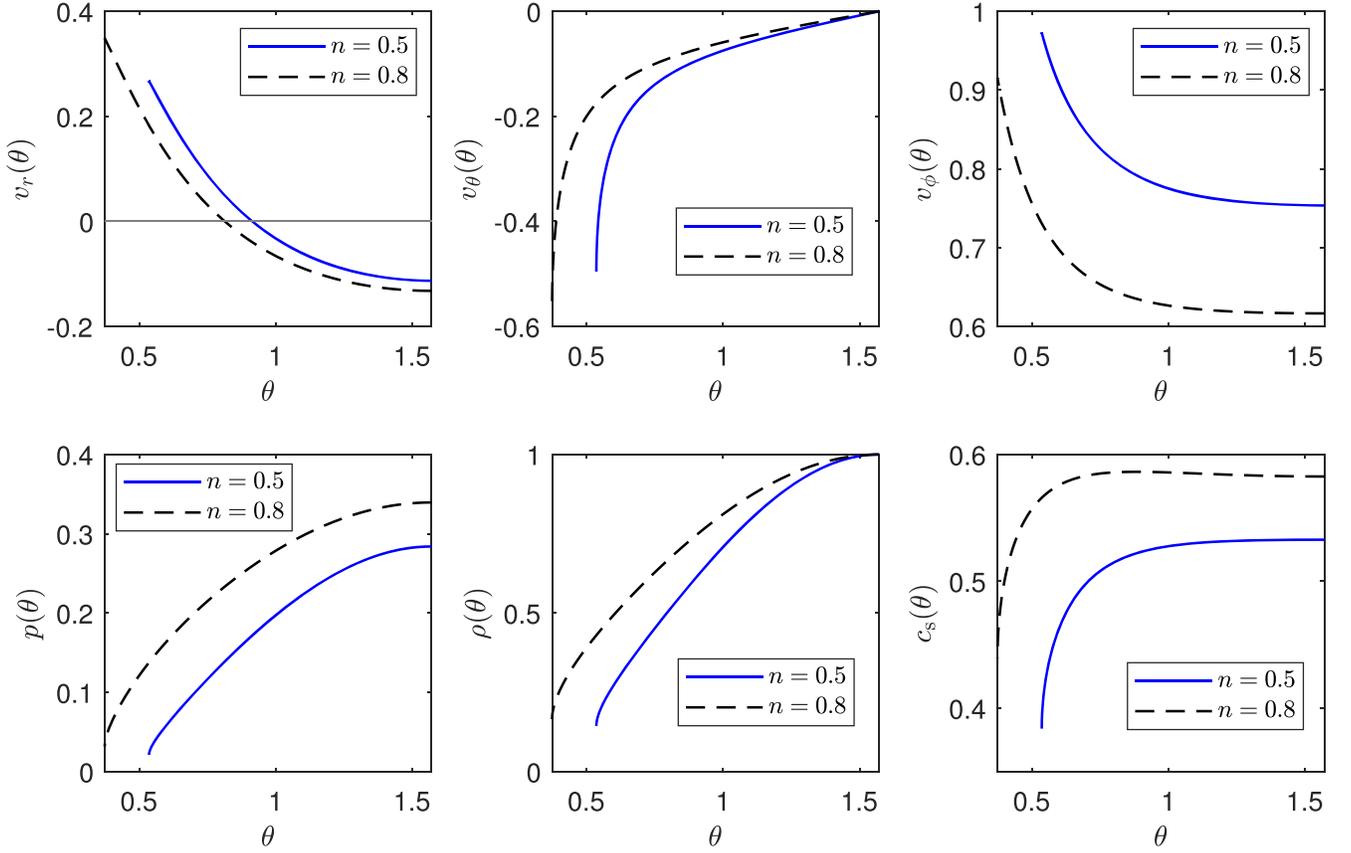

**Figure 1.** Self-similar solutions for $\gamma = 5/3$, $\alpha = 0.1$, $f = 1$, and $n = 0.5$ (blue solid lines) and 0.8 (black dashed lines).

Note that due to axisymmetry and reflection symmetry, we have $v_\theta = 0$ on both the polar axis and the equatorial plane, so the integration of Equation (19) for $\theta$ from 0 to $\pi/2$ gives

$$\frac{\partial \dot{M}_{net}}{\partial r} = 0, \quad (20)$$

where

$$\dot{M}_{net} \equiv \int_0^{\frac{\pi}{2}} 2\pi r^2 \sin\theta \rho v_r d\theta \quad (21)$$

is the net accretion rate (in the upper hemisphere, exactly speaking), which must be constant according to Equation (20). If self-similarity holds for the whole range of $\theta$, we can further get

$$\dot{M}_{net} = 2\pi\sqrt{GM}\, r^{\frac{3}{2}-n} \int_0^{\frac{\pi}{2}} v_r(\theta)\rho(\theta)\sin\theta d\theta = \text{const.} \quad (22)$$

This tells us that a self-similar steady solution applicable to the entire $\theta$ range requires either $n = 3/2$, which will be discussed in Section 4.2, or $\dot{M}_{net} = 0$, which means that the mass outflow exactly cancels the mass inflow (e.g., Xu & Chen 1997; Mosallanezhad et al. 2021; Zeraatgari et al. 2021). The latter case certainly cannot last forever for a real accretion flow, so there must be some non-self-similar structure around the polar axis in the steady state. Our solutions and interpretation also agree quite well with a detailed analysis of numerical simulation data (Yuan et al. 2015), which found that the main body of inflow resides around the equatorial plane, the main body of the wind occupies the $\theta$ range of 30°–60° with an average poloidal speed $v_p\, (=\sqrt{v_r^2 + v_\theta^2})$ of $\sim 0.25\, v_K$, and there exists a much faster disk jet ($\sim 0.3\,c - 0.4\,c$) with much lower density for $\theta \lesssim 15°$. Our self-similar outflow, whose velocity component profiles are proportional to $v_K$, corresponds to the disk wind in Yuan et al. (2015), while the expected non-self-similar structure beyond the outflow corresponds to the disk jet and the transition region between the wind and the jet.

The advective heat transport and its different components in these two solutions are presented in Figure 2, after being normalized with $(GM)^{3/2} r^{-n-5/2}$, in accordance with other self-similar quantities. The top left panel shows the radial advection, and we can see that the advected internal energy in the $r$ direction, $q_{r1}$, becomes 0 for $n = 0.5$, and is positive (cooling) in the inflow and negative (heating) in the outflow for $n = 0.8$, in agreement with the analyses in Section 3. The pressure work in the $r$ direction, $q_{r2}$, is always negative (heating) in the inflow due to the compression of material, and always positive (cooling) in the outflow due to the expansion of material. The total radial advection, $q_r$, as shown in the lower panels, is always a heating mechanism in the inflow and a cooling mechanism in the outflow, as the pressure work dominates over the advected internal energy in the $r$ direction for $n < 3\gamma/2 - 1$.

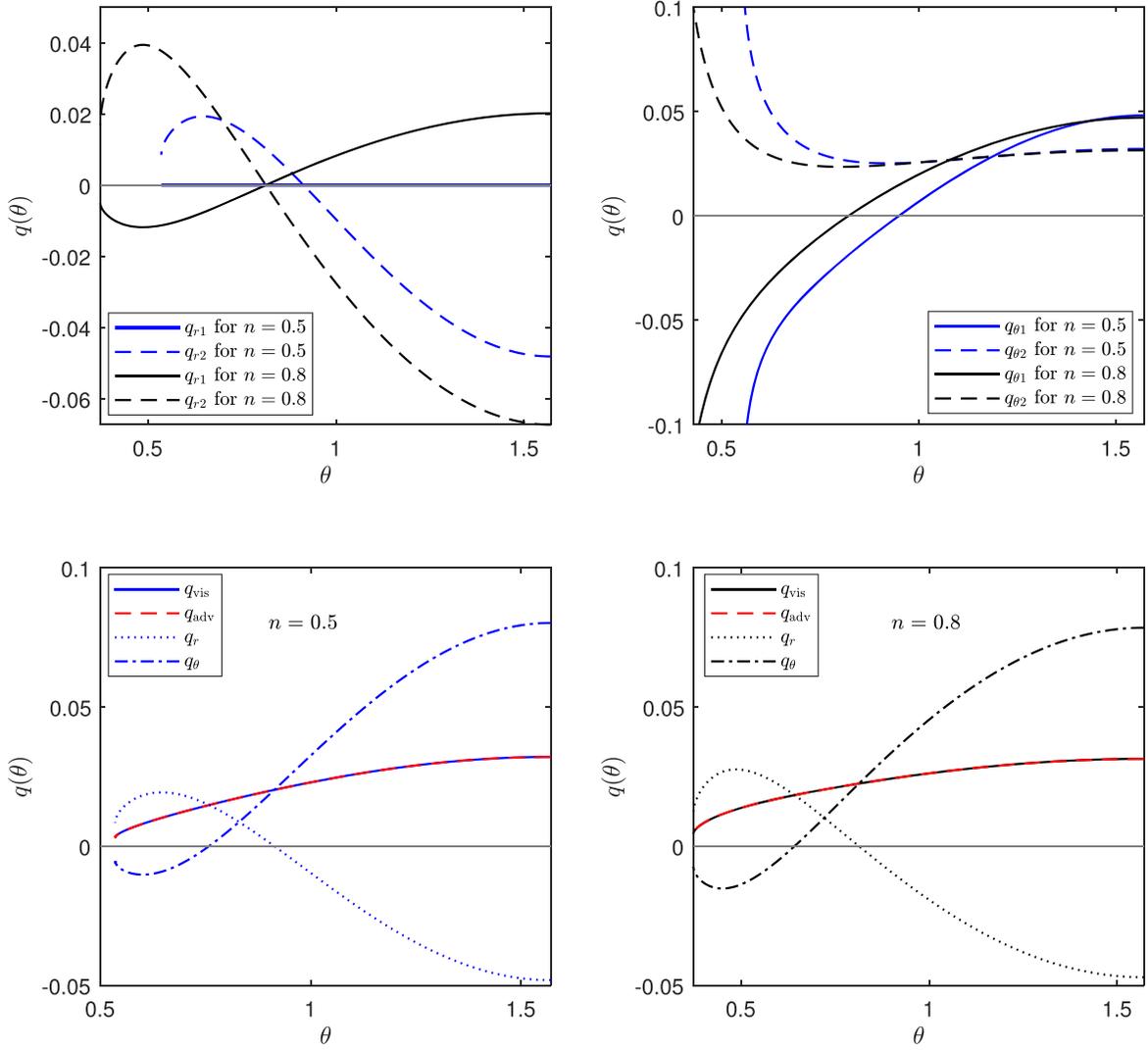

**Figure 2.** Components of advective heat transport as functions of $\theta$ for the solutions of $n = 0.5$ and 0.8, both with $\gamma = 5/3$, $\alpha = 0.1$, and $f = 1$, normalized with $(GM)^{3/2} r^{-n-5/2}$. Positive values represent cooling, and negative values heating. We present the solution of $n = 0.5$ with blue lines, and $n = 0.8$ with black lines. The top left panel shows the radial advection, and the top right panel shows the latitudinal advection, with solid lines for advected internal energy and dashed lines for pressure work. The bottom panels show viscous heating $q_{\rm vis}$ with solid lines, total advection $q_{\rm adv}$ with red dashed lines, total radial advection $q_r$ with dotted lines, and total latitudinal advection $q_\theta$ with dotted–dashed lines, for solutions of $n = 0.5$ on the left and $n = 0.8$ on the right, respectively. The gray solid line in each panel indicates where the function values equal 0.

The latitudinal components of advection are shown in the top right panel of Figure 2. The advected internal energy in the $\theta$ direction, $q_{\theta 1}$, starts out positive (cooling) on the equatorial plane and gradually transforms to a heating mechanism as $\theta$ decreases. Note that the place where $q_{\theta 1} = 0$ does not coincide with the inflow–outflow boundary, and actually $q_{\theta 1}$ has already become a heating mechanism somewhere in the inflow. This effect is offset by the latitudinal pressure work $q_{\theta 2}$, which is always a cooling mechanism, and the total effect of latitudinal advection, $q_\theta$, is always a cooling mechanism in the inflow, which balances the heating caused by viscous dissipation and radial advection. $q_\theta$ remains a cooling mechanism in the outflow for some range of $\theta$ (the inflow–outflow boundary is where $v_r$ and consequently $q_r$, represented by the dotted lines, becomes 0), where the cooling via radial advection is still low, to keep the energy conservation in the steady state. As $\theta$ continues to decrease, the cooling caused by radial advection of outflow becomes large enough, which carries away both the net incoming entropy in the $\theta$ direction from lower latitudes (deducting the entropy going to even higher latitudes) and the entropy generated by local viscous dissipation. This can be perceived as entropy being transferred from the inflow and the low-speed outflow to the high-speed outflow, which is then carried outward. Energy conservation is thus maintained everywhere, as shown by the overlapped solid and dashed lines in the lower panels in Figure 2.

For $\gamma = 5/3$ and $n < 3/2$, we will always see radial advection as a heating mechanism in the inflow. To show a solution with a cooling radial advection in the inflow, we need to reduce the value of $\gamma$. Here we set $\gamma = 4/3$, which actually corresponds to the radiation-pressure-dominated slim disk model. The solutions of $n = 1$ and 1.2 are calculated for comparison, represented by blue solid lines and black dashed lines in Figure 3, respectively. Other parameters are the same as the ADAF solutions, which are $\alpha = 0.1$ and $f = 1$. We can see that the accretion flow still consists of an inflow region, an outflow region, and a non-self-similar region near the polar axis beyond our calculation. The corresponding advective heat transport and its different components are presented in Figure 4.

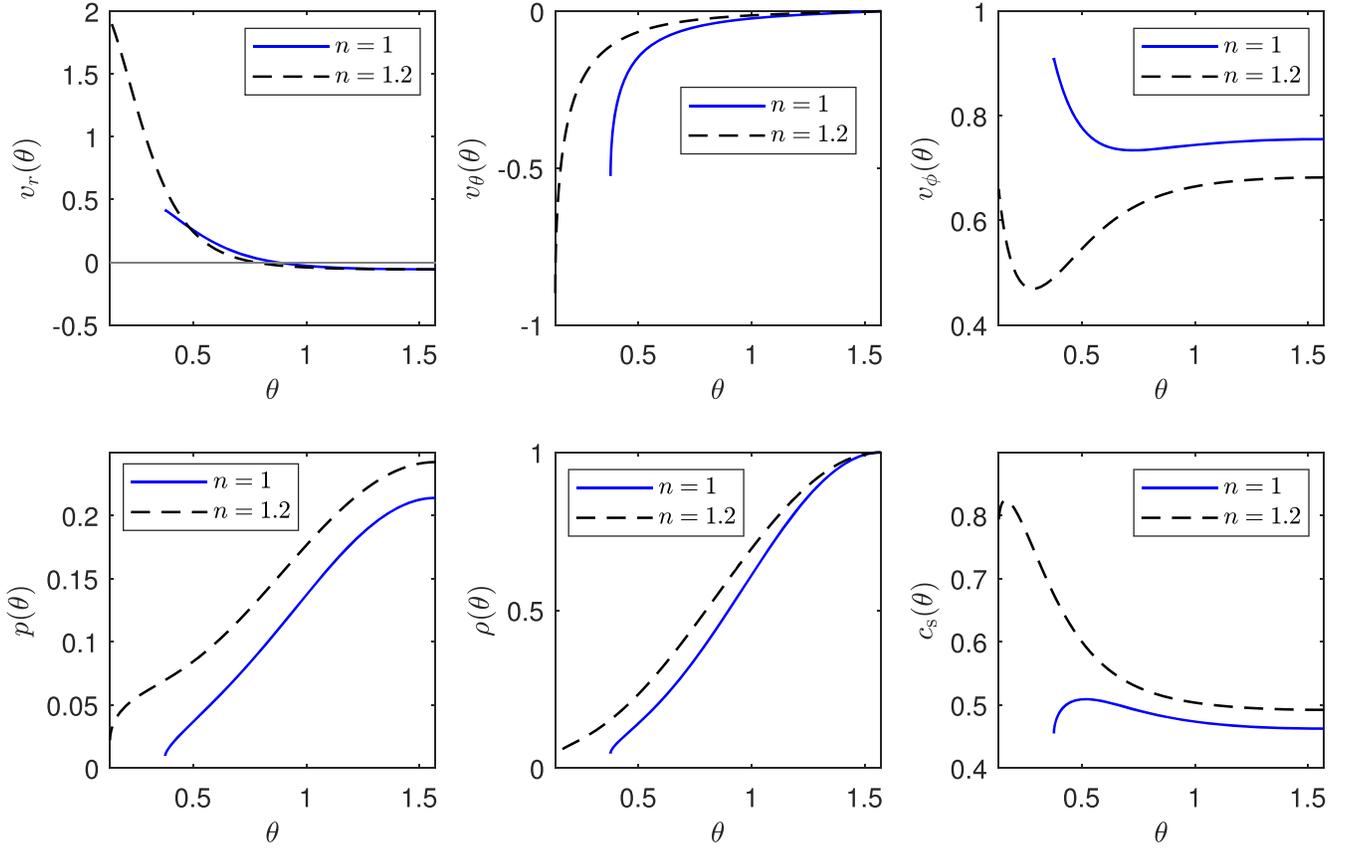

**Figure 3.** Self-similar solutions for $\gamma = 4/3$, $\alpha = 0.1$, $f = 1$, and $n = 1$ (blue solid lines) and 1.2 (black dashed lines).

As expected, the signs of $q_{r1}$ and $q_{r2}$ actually follow the same pattern as the solution of $n = 0.8$. The difference is that, when $n$ is large enough for the smaller value of $\gamma = 4/3$, the radially advected internal energy will dominate over the radial pressure work, as shown in the top left panel of Figure 4. For the solution of $n = 1$, $q_{r1}$ and $q_{r2}$ cancel out each other, and the viscous heating is completely balanced by the latitudinal advection everywhere, so that the solid, dashed, and dotted–dashed lines overlap with each other in the bottom left panel of Figure 4. For the solution of $n = 1.2$, the radial advection becomes a cooling mechanism in the inflow, however, this cooling is not large enough to balance the local viscous heating, and the cooling via latitudinal advection is still required. On the other hand, in the outflow, the radial advection now becomes a heating mechanism, because in the $r$ direction the outflowing internal energy now surpasses the effective cooling caused by expansion of material. To maintain energy balance, this entropy gain is transferred upwards together with the entropy generated by viscous dissipation. Thus for the solutions of $n = 1$ and 1.2, the self-similar outflow does not help to carry away entropy from the inflow, and all this accumulated entropy is eventually transported into the non-self-similar region out of our calculation boundary, which may be eventually carried away by the "disk jet" proposed by Yuan et al. (2015).

### 4.2. The Pure Inflow Case for n = 3/2

There exists an analytical solution for the special case of $n = 3/2$ (Shadmehri 2014; also see Habibi et al. 2017 for an analytical solution with a $\theta$-dependent viscosity coefficient), in the form

$$v_r(\theta) = -\frac{\sqrt{2}\,\alpha}{\sqrt{\alpha^2 + 2(\epsilon')^2 + 5\epsilon'}}, \quad (23)$$

$$v_\theta(\theta) = 0, \quad (24)$$

$$v_\phi(\theta) = \frac{\sqrt{2}\,\epsilon'}{\sqrt{\alpha^2 + 2(\epsilon')^2 + 5\epsilon'}}, \quad (25)$$

$$\rho(\theta) = \rho\left(\frac{\pi}{2}\right)\sin^{\epsilon'}(\theta), \quad (26)$$

$$p(\theta) = \rho\left(\frac{\pi}{2}\right)\sin^{\epsilon'}(\theta)\frac{2\epsilon'}{\alpha^2 + 2(\epsilon')^2 + 5\epsilon'}, \quad (27)$$

where $\epsilon'$ is defined following Narayan & Yi (1995a) as

$$\epsilon' \equiv \frac{1}{f}\left(\frac{5/3 - \gamma}{\gamma - 1}\right). \quad (28)$$

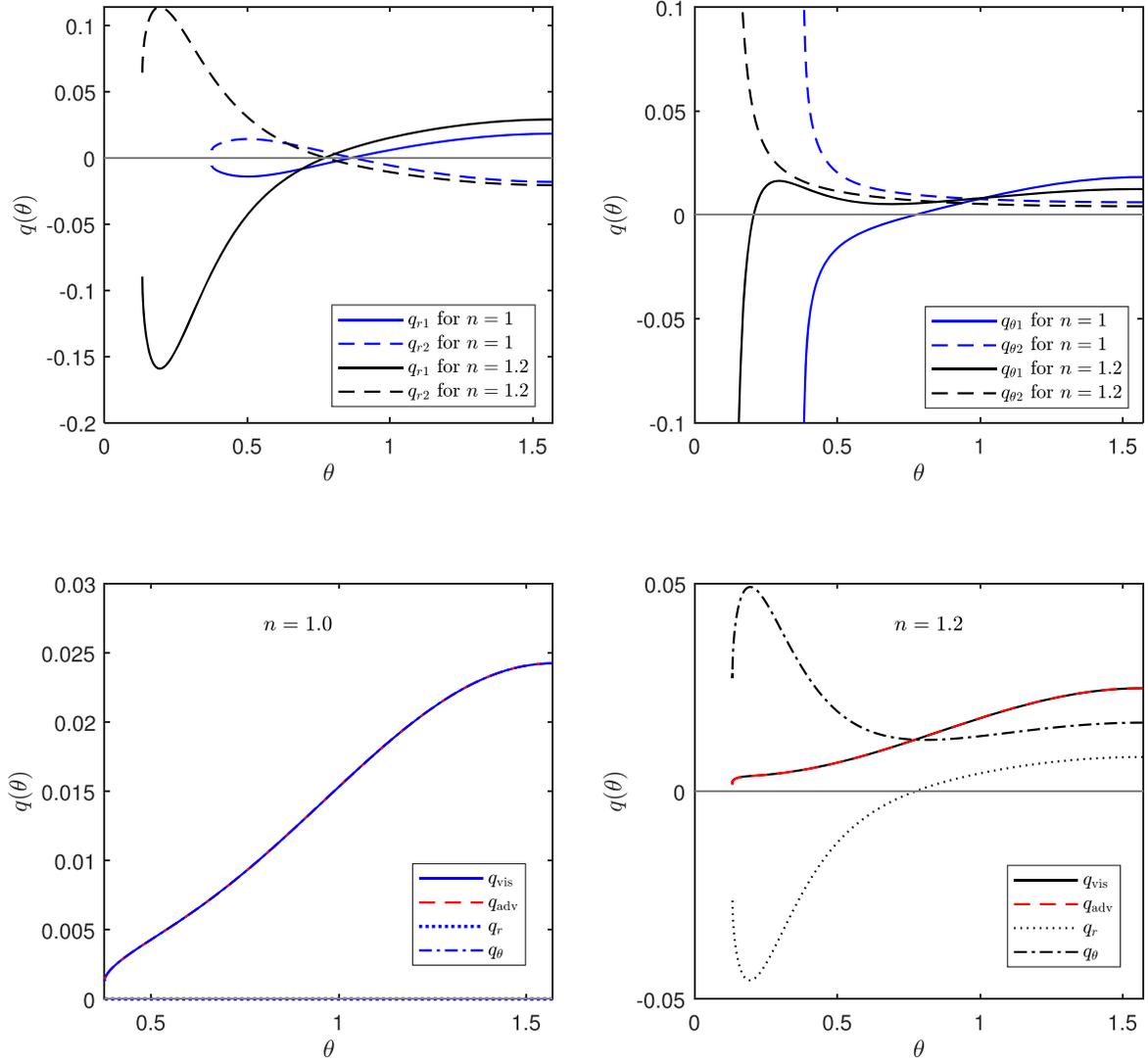

**Figure 4.** Components of advective heat transport as functions of $\theta$ for the solutions of $n = 1$ and 1.2, both with $\gamma = 4/3$, $\alpha = 0.1$, and $f = 1$, normalized with $(GM)^{3/2} r^{-n-5/2}$. Positive values represent cooling, and negative values heating. We present the solution of $n = 1$ with blue lines, and $n = 1.2$ with black lines. The top left panel shows the radial advection, and the top right panel shows the latitudinal advection, with solid lines for advected internal energy and dashed lines for pressure work. The bottom panels show viscous heating $q_{\mathrm{vis}}$ with solid lines, total advection $q_{\mathrm{adv}}$ with red dashed lines, total radial advection $q_r$ with dotted lines, and total latitudinal advection $q_\theta$ with dotted–dashed lines, for solutions of $n = 1$ on the left and $n = 1.2$ on the right, respectively. The gray solid line in each panel indicates where the function values equal 0.

Note that $v_r(\theta)$, $v_\phi(\theta)$, and $p(\theta)/\rho(\theta)$ are all constants and $v_\theta = 0$ in the solution, which is more simplified than the numerical solution of Narayan & Yi (1995a), which adopts the full viscous stress tensor instead of only the $r\phi$ component and the boundary conditions on the polar axis where $v_r(\theta)$ is set to 0. The accretion flow consists of pure inflow, and the viscous heating is balanced solely by the radial advection. This requires that the radially advected internal energy dominates over the radial pressure work which is always a heating mechanism in the inflow. When we set $\gamma = 5/3$, they can only cancel out each other for $n = 3/2$ according to Equations (14) and (15), so that the viscous heating should also be 0 to maintain energy conservation, which in turn causes $v_\phi$ to become 0 and the solution basically represents freefall of accreted material without rotation and viscous dissipation. Another concern is that, to avoid singularity of the angular velocity $\Omega = v_\phi/(r \sin \theta)$, $v_\phi$ should be 0 on the polar axis, so the solution should stop before reaching there, except for the freefall case. The solution of Narayan & Yi (1995a) also has this problem (their $\Omega[\theta]$ in Figure 1 is our $v_\phi[\theta]$, just with a different notation), and they suggest that bipolar outflow should be developed.

Gu (2015) used the polytropic relation, $p = K\rho^\Gamma$, in place of the energy equation and studied the energy balance in ADAFs for $n = 3/2$ under self-similar assumptions. He proposed that advective cooling is not enough to balance viscous heating in this case and outflow should be inevitable. His work was revisited by Zahra Zeraatgari & Abbassi (2015) with a $\theta$-dependent kinematic viscosity coefficient, and by Zahra Zeraatgari et al. (2018) in which a large-scale toroidal magnetic field is further added. Both works proposed that energy balance could be achieved for certain values of $\theta$, and the $\theta$-integrated heating and cooling could also balance each other for certain selection of parameters. The analytical solution here, however, shows that energy balance could actually be achieved simultaneously for all $\theta$ values with an isothermal ($\Gamma = 1$) latitudinal structure.

Here we emphasize that, as mentioned in Section 3, the range of $n$ is likely 0.5–1 for ADAFs, so the pure inflow solution discussed here is mainly of academic interest rather than applicable to real accretion flows.

## 5. Summary and Conclusion

Advective heat transport is an effective heating/cooling mechanism that represents the net entropy difference between the incoming and the outgoing material for a fixed region in the rest frame, which obviously depends on the direction of motion. To study the difference of advection in the inflow and the outflow that have opposite radial motion, we divide the advection into two parts, the radial advection and the latitudinal advection in spherical coordinates, each of which are then further divided into the advected internal energy term and the pressure-work term. We then analyze their profiles with self-similar solutions of ADAFs. We find that, for inflow–outflow solutions with $n < 3/2$, the radial pressure work $q_{r2}$ is always a heating mechanism in the inflow due to compression, and a cooling mechanism in the outflow due to expansion of material. The radially advected internal energy $q_{r1}$ acts similarly with $q_{r2}$ for $n < 1/2$, becomes 0 for $n = 1/2$, and acts oppositely for $n > 1/2$. The total radial advection is a heating mechanism in the inflow and a cooling mechanism in the outflow for $n < 3\gamma/2 - 1$, becomes 0 for $n = 3\gamma/2 - 1$, and turns into cooling in the inflow and heating in the outflow for $n > 3\gamma/2 - 1$. The latitudinally advected internal energy generally starts out as a cooling mechanism on the equator and gradually transforms to a heating mechanism as $\theta$ decreases. The latitudinal pressure work is always a cooling mechanism in our numerical solutions. The total latitudinal advection $q_\theta$ takes the form that maintains energy conservation.

An analytical solution exists for the special case of $n = 3/2$, which consists of pure inflow with $v_\theta = 0$. Consequently there is no latitudinal advection, and viscous heating is balanced by radial advection alone, which is always a cooling mechanism because the advection of internal energy dominates over the pressure work. In this case $v_r$ and $v_\phi$ are independent of $\theta$, and the latitudinal structure is isothermal.

Theoretical and numerical simulation works in literature indicate that $n \in [0.5, 1]$. In this range, radial advection is actually a heating mechanism in the inflow, contrary to the popular interpretation of "advective cooling," because the radial pressure work dominates over the radial advection of internal energy. The steady state can thus only be achieved when latitudinal advection is considered.

Our study confirms and quantifies the perception that outflow (or wind) carries away entropy thus cooling the inflow (e.g., Blandford & Begelman 1999, 2004; Xie & Yuan 2008; Begelman 2012), with the latitudinal advection $q_\theta$ that represents the net entropy loss (or gain for $q_\theta < 0$) for a fixed region due to latitudinal motion. However, the outflow itself must first have enough ability to carry away entropy radially to overcome the local viscous heating, or the incoming entropy from lower latitudes cannot be carried away and is transferred to even higher latitudes instead. For $n < 3\gamma/2 - 1$, this happens in the outflow just beyond the inflow–outflow boundary, where the ability of the outflow to carry away entropy is still low due to low radial velocity. For $n > 3\gamma/2 - 1$, the radial advection of outflow is actually a heating mechanism, so entropy cannot be carried away radially at all, and keeps going upward and accumulating, which eventually goes out of our calculation boundary and is carried away by the non-self-similar structure beyond, possibly the "disk jet" reported by Yuan et al. (2015).

The numerical results in this paper are based on self-similar assumptions that have been proposed to behave well in a range of intermediate radii. We think that the conclusions here should apply to more general cases, although exact analyses in the global two-dimensional solution (e.g., Kumar & Gu 2018) and numerical simulations still remain to be done.

The analysis in this paper does not apply to CDAFs, in which the energy balance is maintained mostly by convective energy fluxes. However, some recent numerical simulations found that ADAFs are convectively stable (Yuan et al. 2012; Narayan et al. 2012), and study of trajectories of the virtual Lagrangian particles found that strong "real" outflow does exist in ADAFs (Yuan et al. 2015; Yang et al. 2021), so we think that convective energy fluxes can be safely ignored.

We assume that only the $r\phi$-component of the viscous stress is nonzero, which is a caveat in our work. On the other hand, in the energy equation, the inclusion of other components of the viscous stress does not change the formula of advective heat transport, and only changes the formula of the viscous heating, which would still be a heating term anyway, so the analysis in Section 3 still holds. It does affect the expanded form of the equation of motion, and we are planning to address this caveat in our future work.

We thank the referee for helpful comments. This work is supported by the Natural Science Foundation of China (grant 11703083).

## ORCID iDs

Cheng-Liang Jiao (焦承 亮)  https://orcid.org/0000-0002-7663-7900